# Inventory competition in a multi channel distribution system: The Nash and Stackelberg game

Mahtab Hoseininia                                      M_hosseininia@cic.aut.ac.ir

*Department of Industrial engineering, Amirkabir University of Technology, Tehran, Iran.*

Farzad Didehvar                                        Didehvar@cic.aut.ac.ir

*Department of mathematic & computer science, Amirkabir University of Technology, Tehran, Iran.*

Mir Mehdi Seyyed Esfahani                              Msesfahani@cic.aut.ac.ir

*Department of Industrial engineering, Amirkabir University of Technology, Tehran, Iran.*

**Abstract.** This paper investigates inventory management in a multi channel distribution system consisting of one manufacturer and an arbitrary number of retailers that face stochastic demand. Existence of the pure Nash equilibrium is proved and parameter restriction which implies uniqueness of it is derived. Also the Stackelberg game where the manufacturer plays a roll as a leader is discussed. Under specified parameter restrictions which guarantee profitability, sufficient condition for uniqueness of Stackelberg equilibrium is obtained. In addition comparison with simultaneous move game is made. The result shows that when whole prices are equal to production cost, manufacturer carries more inventory than simultaneous move game.

## 1. Introduction

For any company with a product to sell, making that product available to the intended customers can be as crucial a strategic issue as developing the product itself. While distribution strategy is a very traditional concern, for many companies it has recently come under intense scrutiny due to a number of major developments. First, the expanding role of the Internet in consumer and business procurement activity has created unprecedented opportunities for easy and vast access to customers. Furthermore, the economics of materials delivery has been revolutionized by the efficient and pervasive logistical networks deployed by third-party shipping powerhouses, such as Federal Express and United Parcel Services. As a result many manufacturers are reconsidering whether they should rely on intermediaries, sell directly to end customers or even pursue both approaches in parallel (Tsay and Agrawal, 2004).

Multi channel distribution system chooses the third approach for distribution of the products. It consists of a manufacturer and retailers in which a manufacturer sells its products through independent retailers as well as through her wholly-owned channel. Therefore the manufacturer simultaneously acts as a supplier as well as competitor to the retailers.

This type of distribution allows firms to extend their market coverage by employing various distribution channels because firms can target many different customer segments and reach new customer segments more efficiently. Moreover, the use of multi channel distribution enhances the ability to meet the needs of existing customers. Since purchase experiences vary among different channels, so customers have different preferences with respect to purchase experiences. Price and non-price factors such as location, product assortment and customer service also influence customers' channel choices. In addition, more extensive market presence cause to two subjects: increase the customers' awareness and establishment of higher brand loyalty for existing as well as future products.

In a multi channel distribution system the manufacturer and retailers present substitutable product. Therefore when a demand can not be satisfied by each of them, because of stock out, the customer may go to other channels. So the profit function of channels is influenced not only by her own order decision but also by the decisions of her competitors. Thereby a strategic interaction among channel inventory decision is created and game theoretic is applied to analyze it.

In this paper a game theoretic inventory management model for multi channel distribution which consists of one manufacturer and an arbitrary number of retailers is developed over an infinite horizon. Manufacturer and retailers face stochastic demand. They use base stock policy to control inventory. We focus on equilibrium strategies of manufacturer's and retailers' channels while players participate the Nash and Stackelberg game. Also comparison between these equilibriums is made.

The rest of the paper is organized as follows: Section 2 presents the literature review on inventory competition in supply chain. The main model is presented in Section 3. Section 4 discusses inventory management in a multi channel distribution in the two frameworks of the Nash and Stackelberg games. Section 5 summarizes and concluded the paper.

## 2. Literature review

Inventory competition in supply chain is originated from demand substitution. Parlar (1988) develops a game theory model for single period inventory control of two substitutable products as a two person game. He proves existence and uniqueness of the Nash solution. Wang and Parlar (1994) extend Parlar's model (1988) to a three person game and analyze it in both noncooperative and cooperative cases. In noncooperative scenario they prove existence of the Nash equilibrium. In cooperative scenario if side payments are not allowed, they prove the Nash strategies always exist in any case of cooperation and if side payments are allowed, those conditions which imply on the non-emptiness of the game core are investigated. Lippman and McCardle (1997) extend Parlar's model (1988) to an arbitrary number of retailers whereas aggregate industry demand is allocated among firms by splitting rules. They examine the relationship between the Nash equilibrium inventory levels and the demand splitting rules and provide conditions under which there is a unique equilibrium. Mahajan and van Ryzin (2001) analyze inventory management model consists of N competing firms that provide substitutable goods. Demand is stochastic sequence of heterogeneous customers who choose dynamically from the available goods. Existence of the Nash equilibrium is proved and it is shown that under symmetric conditions (identical parameters), this equilibrium is unique. Anupindi et al. (2001) develop a general framework for analysis of inventory management of decentralized distribution system. This framework entails N retailers who hold stocks locally or/and at one or more central warehouses. This consists of two stages. In the first stage in order to satisfy local demand, inventory decisions are made. In the second stage by giving inventory levels and realized demands, allocation of stocks for satisfying residual demands is determined as well as financial decision of allocation of revenues. In first stage sufficient conditions for existence of the pure Nash equilibrium are derived. In the second stage, sufficient conditions for existence of the core for allocation decisions are investigated. Granot and Sošić (2003) extend the model which is proposed by Anupindi et al. (2001). The difference between these two models is about second stage; before shipment of residual supply, each retailer decides how much of her residual supply she wants to share with the other. Avşar and Baykal-Gürsoy (2002) extend Parlar's model (1988) into infinite horizon and lost sale case. Under the discounted payoff criterion and with stationary base stock strategies for inventory control system, existences and uniqueness of the Nash equilibrium is proved. Netessine and Rudi (2003) investigate inventory management of an arbitrary number of substitutable products under both centralized inventory management and competition. They show that for noncompetitive case objective function is not necessarily concave function. Also a necessary optimality condition which may not be sufficient is obtained. For competitive case existence of the Nash equilibrium is proved and conditions that imply uniqueness are obtained. Also optimally conditions for the competitive N case are obtained and it is compared with noncompetitive solution. Dai et al. (2005) analyze inventory control decisions of two retailers in a distribution system that competes for both supply capacity and customers. The necessary and sufficient conditions for existence of a unique Nash equilibrium are derived. In case the Nash equilibrium does not exist, the concept of the Stackelberg game is used to develop optimal strategies for both the leader and the follower. Boyaci (2005) presents an inventory management model in a multi channel distribution system over an infinite horizon and proves existence and uniqueness of the Nash equilibrium. Netessine et al. (2006) present inventory control model for two retailers in multi period environment whereas various customer responses to a retailer's backorder is considered. Serin (2007) based on newsvendor problem, discusses Nash equilibrium and Stackelberg equilibrium of two retailers and present conditions which imply leader's profit doesn't improve from a simultaneous game. They show that under appropriate conditions stationary base stock inventory policy is the Nash equilibrium of the game. Geng and Malik (2007) based on newsvendor problem develop an inventory control model for multi channel distribution and propose a dynamic game where the retailer is Stackelberg leader and the manufacturer is Stackelberg follower who has the authority to undercut retailer's order. Both capacitated and infinite capacity games are discussed and equilibrium of the games is derived. Yao et al. (2009) discuss three inventory strategies in a multi channel distribution system. These strategies are: centralized inventory strategy, a Stackelberg inventory strategy whereas the manufacturer plays a roll as a leader and a strategy where the manufacturer outsources to a 3PL. They obtain and compare optimal inventory levels. Chiang (2009) develop Markov inventory management model for multi channel distribution and discusses the Nash and Stackelberg games. Also channel conflict induces from simultaneously vertical and horizontal competition is discussed. McGillivray and Silver (1978), Parlar and Goyal (1984) also Ernst and Kouvelis (1999) discuss inventory competition in a single firm context.

## 3. Modeling assumptions and notations

We consider a manufacturer who produces a single product at a unit cost $c \geq 0$. She distributes it through her wholly owned sales channel at a price $p_m$ and through an independent retailer i sales channel $(i=1,...,n)$ at a wholesale price $w_{r_i}$. The retailers will resell the product through their own channels at a price $p_{r_i}$. All prices are exogenous. In order to avoid trivial solutions it is assumed that:

$$c < w_{r_i} < p_{r_i} \ (i=1,...,n) \text{ and } c < p_m$$

This set of assumptions guarantees that it is profitable to sell the product in all of the channels.

Each channel has a limited local monopoly as a result of some form of channel differentiation (in terms of location, brand name, prices, service and support provided, etc). Therefore the total market demand is shared between channels. Each channel has customers referring to it as their first choice for satisfying their demand. This is called primary demand. We will let $\tilde{D}_j \ (j = M, R_1, ..., R_n)$ denote the primary demands of channel j (M indicates the manufacturer channel and $R_i$ indicates retailer i channel). It is assumed that the demand profile $(\tilde{D}_m, \tilde{D}_{r_1}, ..., \tilde{D}_{r_n})$ follows a known, continuous joint distribution, with a corresponding set of marginal strictly increasing distribution function. For analytical tractability, we suppose that the demands across different periods are independent and identically distributed.

If one channel is out of stock, a fraction of unsatisfied customers visiting the other channels. The substitutability of the channels product is modeled through substitution rates (also it is called market search rates) $\alpha_{jk}$ which are known.

$\alpha_{jk}$ $(j, k = M, R_1,...,R_n, j \neq k)$ is a fraction of channel j demand that switch to channel k when channel j is out of stock $(0 \leq \alpha_{jk} \leq 1, \sum_j \alpha_{jk} \leq 1)$. Substitution between channels occurs only one time. Demand substitution by any other products is not allowed. In the event that the product is out of stock at both channels, we assume that the demand will be lost. It is considered that there are no "after effects" of lost customer demands.

In this model, it is assumed that replenishment lead times are zero and the manufacturer has infinite supply capacity at the upper echelon, which means that he can instantaneously supply the orders of manufacturer's and retailers' channels. The manufacturer does not hold any stock at this level; stocks are kept by the manufacturer and the retailers only to satisfy end customer demands. (This is a common assumption in models with simultaneous horizontal and vertical channel interactions).

The stocking decisions in the supply chain are made periodically and the inventory is controlled according to base stock policies. The sequence of events which explain the dynamics of the model is as followed: At the beginning of each period, the on-hand inventory is observed and replenishment orders are given accordingly. These orders are received immediately under the infinite supply capacity and zero lead times assumptions. Then, the channel demands are fulfilled, unmet demand is lost, and excess stocks are carried over to the next period. $S_j$ and $h_j$ indicate the base stock level and holding cost of channel j respectively.

As indicated by Anupindi and Bassok (1999) also Boyaci (2005), the above dynamic model, under an undiscounted, infinite-horizon and average profit maximization criterion can be reduced to an equivalent single-period static (newsvendor-type) model. In the remainder of the paper we use this equivalent single period model.

The total composite demand for each channel denoted by $D_j$ includes first choice customers as well as any spill over customers substituting the product. It is given by:

$$D_j = \tilde{D}_j + \sum_k \alpha_{kj}(\tilde{D}_k - S_k)^+ \quad (j, k = M, R_1,...,R_n, j \neq k) \tag{1}$$

Therefore, in spite of independency of the first choice demand at all of the channels, the actual demand faced by each channel depends on the respective first choice demand and stock level of other channels.

Since the manufacturer and the retailers don't undertake any penalty cost, cost structure only consist of production (order) and inventory holding costs. So the manufacturer's expected profit function is given by:

$$\Pi_m(S_m, S_{r_1},...,S_{r_n}) = (p_m - c)S_m - (p_m - c + h_m)E(S_m - D_m)^+ + \sum_{i=1}^{n}(w_{r_i} - c)S_{r_i} - \sum_{i=1}^{n}(w_{r_i} - c)E(S_{r_i} - D_{r_i})^+ \quad (2)$$

And retailer i by:

$$\Pi_{r_i}(S_m,...,S_{r_i},...,S_{r_n}) = (p_{r_i} - w_{r_i})S_{r_i} - (p_{r_i} - w_{r_i} + h_{r_i})E(S_{r_i} - D_{r_i})^+ \quad (3)$$

A schematic representation of a multi channel distribution system is shown in Fig.1.

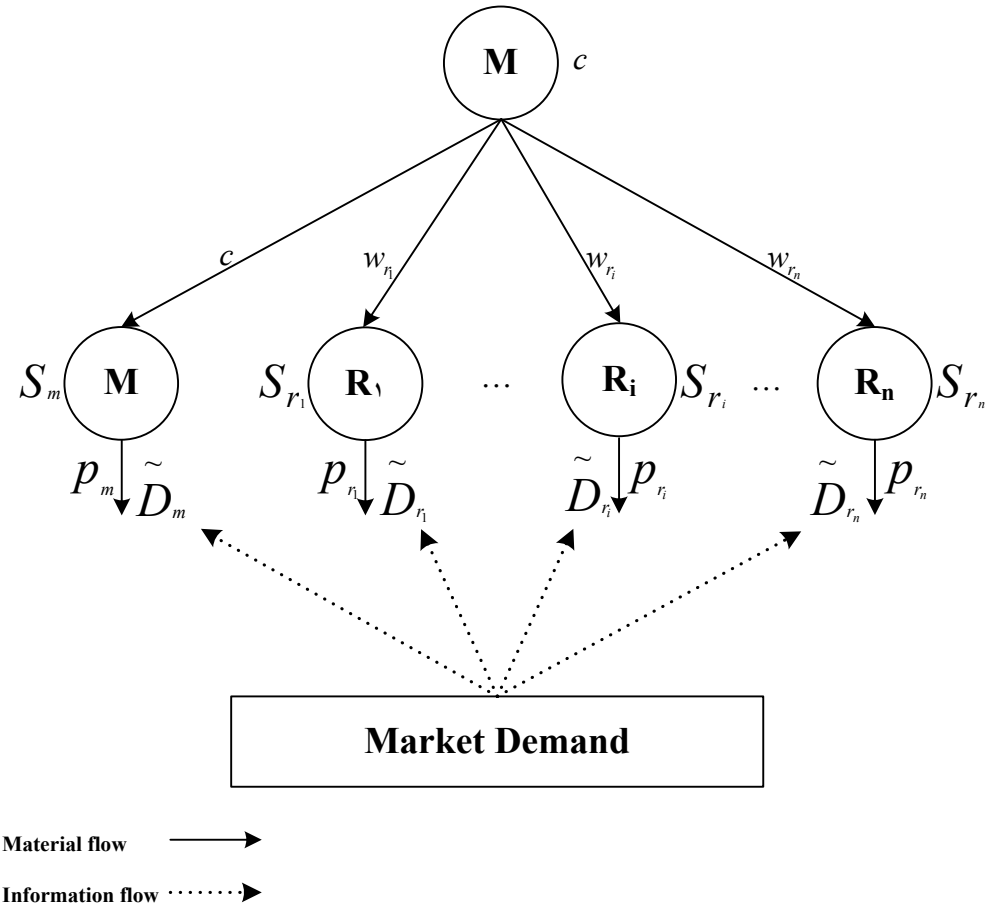

Fig. 1. Manufacturer - n retailers multi channel distribution system.

# 4. Analysis

## 4.1. The Nash game and the Nash equilibrium

In the Nash game there isn't any dominance relation between the supply chain members and the manufacturer and retailers simultaneously choose their base stock levels with the objective of maximizing their respective profits. So it is a one shot game: The players choose the base stock levels at time 0 and maintain them over an infinite horizon.

The Nash equilibrium predicts rational behavior of players; no player wants to unilaterally deviate from it since such behavior would lead to lower payoffs.

**Lemma 1.** Assume the following holds.

*Condition C1.* $(p_m - c + h_m) > \sum_{i=1}^{n} \alpha_{mr_i}(w_{r_i} - c)$. Then:

   a. $\Pi_m(S_m, S_{r_1},...,S_{r_n})$ is concave in $S_m$ for fixed $(S_{r_1},...,S_{r_n})$, and is submodular in $(S_m, S_{r_1},...,S_{r_n})$.

   b. $\Pi_{r_i}(S_m,...,S_{r_i},...,S_{r_n})$ is concave in $S_{r_i}$ for fixed $(S_m,...,S_{r_{i-1}}, S_{r_{i+1}},...,S_{r_n})$, and is submodular in $(S_m, S_{r_1},...,S_{r_n})$.

Condition C1 summarizes the parameter restriction that is necessary to avoid trivial solutions and to guarantee the concavity of manufacturer profit function. This inequality implies that it is more profitable for manufacturer to satisfy first choice demand rather than denying the sale and forcing the customer to substitute.

Furthermore, submodularity of manufacturer and retailers profit functions imply that product of channels are economic substituent; increasing stocking level in one channel results in a decreasing marginal profit of increasing stocking level of the other channels.

**Lemma 2.** The manufacturer channel has a unique best response base stock level $S_m(S_{r_1},...,S_{r_n})$ given by:

$$S_m(S_{r_1},...,S_{r_n}) = \left\{ S_m : P(D_m \leq S_m) = \frac{p_m - c - \sum_{i=1}^{n} \alpha_{mr_i}(w_{r_i} - c)P(D_{r_i} \leq S_{r_i}, \tilde{D}_m > S_m)}{p_m - c - h_m} \right\} \quad (4)$$

Retailer i has a unique best response base stock level $S_{r_i}(S_m,...,S_{r_{i-1}}, S_{r_{i+1}},...,S_{r_n})$ given by:

$$S_{r_i}(S_m,...,S_{r_{i-1}},S_{r_{i+1}},...,S_{r_n}) = \left\{ S_{r_i} : P(D_{r_i} \leq S_{r_i}) = \frac{p_{r_i} - w_{r_i}}{p_{r_i} - w_{r_i} - h_{r_i}} \right\} \quad (5)$$

It is important to note that the assumption $w_{r_i} < p_{r_i}$ results in stocking the product in equilibrium by the retailers. However, there might be situation in which it is more profitable for manufacturer not to sell directly to customers and forcing them to substitute the product in the retailers' channels.

Assume the following holds.

*Condition C2.* $p_m - c > \sum_{i=1}^{n} \alpha_{mr_i}(w_{r_i} - c)$

Under condition C2 the manufacturer channel always stock the product in equilibrium. This condition implies that manufacturer's profit margin from customers in her own channel is greater than marginal profit she would be gained from forcing the customers to substitute the product in the retailers' channels. If this condition is fulfilled, the equilibrium is interior point.

**Proposition 1:** If condition C1 is satisfied, inventory management game in a multi channel distribution system admits at least one pure Nash equilibrium which is obtained as follow:

$$\begin{cases} P(D_m \leq S_m^*) = \dfrac{p_m - c - \sum_{i=1}^{n} \alpha_{mr_i}(w_{r_i} - c) P(D_{r_i} \leq S_{r_i}^*, \tilde{D}_m > S_m^*)}{p_m - c + h_m} \\ \\ P(D_{r_i} \leq S_{r_i}^*) = \dfrac{p_{r_i} - w_{r_i}}{p_{r_i} - w_{r_i} + h_{r_i}} \quad (i=1,...,n) \end{cases} \quad (6)$$

**Proposition 2:** If condition C2 is satisfied, the Nash equilibrium of inventory management game in a multi channel distribution system is unique.

### 4.2. Stackelberg game and Stackelberg equilibrium

The manufacturer in multi channel distribution system simultaneously acts as a supplier and competitor . Therefore she has a full control on her own channel. So in this section we discuss the Stackelberg game whereas the manufacturer acts as a Stackelberg leader, chooses her base stock level first and the retailers are Stackelberg followers and respond to any decision of the manufacturer.

In this paper backward induction is used to characterize equilibrium of a Stackelberg game. Since the retailers act as a follower, first best response of retailer i, $S_{r_i}^{**}$, to any decision of the manufacturer and the other retailers is obtained. Then the manufacturer's optimal base stock level is derived. Therefore the manufacturer chooses the best possible point on the retailer's response function and at least gain as Nash equilibrium. Since the profit function of manufacturer and retailers are continuous, the Stackelberg equilibrium exists (Cachon and Netessine, 2004). In this section we discuss important features of Stackelberg equilibrium, restrictions on parameters and demand distribution which implies uniqueness of it.

**Lemma 3.** If the base stock level of the manufacturer channel increases, the base stock level of retailer i decreases and the rate of changes in retailers' channels are less than substitution rate from manufacturer channel.

**Lemma 4.** Assume the following holds.

*Condition C3.* $(w_{r_i} - c) > \alpha_{r_i m}(p_m - c + h_m) + \sum_{\substack{l=1 \\ l \neq i}}^{n} \alpha_{r_i r_l}(w_{r_l} - c)\ (i = 1,...,n)$. Then sufficient condition to hold $\dfrac{d}{dS_m}\left(\dfrac{\partial \Pi_m}{\partial S_{r_i}} \dfrac{dS_{r_i}^{**}}{dS_m}\right) \leq 0$ is as follow:

$$\alpha_{mr_i}^2 f'_{D_{r_i}|\tilde{D}_m > S_m}(S_{r_i}) f_{\tilde{D}_{r_i}}(S_{r_i}) P(\tilde{D}_m \leq S_m) P(\tilde{D}_m > S_m) P(D_{r_i} > S_{r_i}) + \qquad (7)$$
$$\alpha_{mr_i}^2 f^2_{D_{r_i}|\tilde{D}_m > S_m}(S_{r_i}) P^2(\tilde{D}_m > S_m) f_{D_{r_i}}(S_{r_i}) - \alpha_{mr_i} f_{\tilde{D}_{r_i}}(S_{r_i}) f_{\tilde{D}_m}(S_m) f_{D_{r_i}}(S_{r_i}) P(D_{r_i} > S_{r_i}) \geq 0$$

Condition C3 implies that profit margin of selling to retailer i for manufacturer is greater than forcing the customers of that channel to substitute the product from manufacturer's channel or other retailers' channels

**Proposition 3:** If sufficient condition of Lemma 4 is fulfilled and parameter restrictions of condition C3 and C1 are satisfied, Stackelberg equilibrium in multi channel distribution while manufacturer plays a role as a leader, is unique.

**Proposition 4:** If $w_{r_i} = c\ (i = 1,...,n)$, manufacturer carries more inventory than a simultaneous game.

While $w_{r_i} = c\ (i = 1,...,n)$, manufacturer obtain no profit of selling product to the retailers and so the multi channel distribution system reduces to N-horizontal competition product. Mahajan and Van Ryzin (2001) proved that channels over stock the product. This theorem shows that when manufacturer plays a roll as a leader tendency to stock the product for manufacturer increases.

## 5. Conclusion and future work

In this paper we have developed inventory management model for multi channel distribution system where it consists of one manufacturer and an arbitrary number of retailers. We proved that the pure Nash equilibrium exists and parameter restriction which implies uniqueness of it is derived. Also we discussed the Stackelberg game whereas the manufacturer play a roll as a leader and select the base stock level first. Necessary condition for uniqueness of the Stackelberg equilibrium is derived and comparison with the Nash equilibrium is done. Result shows that when whole prices are equal to production cost, manufacturer carries more inventory than simultaneous move game.

Our research can be extended in several directions by relaxing the assumption of this paper. We have assumed that the manufacturer has infinite supply capacity. By using queue approach or adding a supply constraint, this assumption can be relaxed. Although presence of manufacturer in distribution system increase market share, but also increase competition in the market. We propose developing a model which identify when it is profitable for manufacturer to engage in distribution system. We have assumed that all of the channels store the product solely to satisfy demand of their respective channels. It will be interesting to extend this model to situation where the manufacturer plays a roll as a upstream member of supply chain and investigate the impact of stocking level of her on downstream members. It is assumed that if demand of channels is unsatisfied, it will be lost. In future work different customer behavior can be considered and impact of the behavior on equilibrium can be investigated.

## Appendices

### Appendix A: Nashequilibrium

**Proof of Lemma 1.** In order to prove the concavity of manufacturer profit function, it is sufficient to show that second partial derivative of $\Pi_m(S_m, S_{r_1}, ..., S_{r_n})$ with respect to its own decision variable $S_m$ is negative.

$$\frac{\partial^2 \Pi_m(S_m, S_{r_1}, ..., S_{r_n})}{\partial S_m^2} = -(p_m - c + h_m)f_{D_m}(S_m) - \sum_{i=1}^{n}(w_{r_i} - c)\frac{\partial^2 E(S_{r_i} - D_{r_i})^+}{\partial S_m^2} \quad (8)$$

The first term is obtained through Leibnitz's rule and the second term is achieved as follow: Based on Netessine and Rudi (2003) since the function under the expectation is integrable and has a bounded derivative, it satisfies the Lipchitz condition of order and hence the expectation and the derivative can be interchanged. So:

$$\frac{\partial E(S_{r_i} - D_{r_i})^+}{\partial S_m} = E(\frac{\partial (S_{r_i} - D_{r_i})^+}{\partial D_{r_i}} \frac{\partial D_{r_i}}{\partial S_m}) = E\left(I_{\{D_{r_i} \leq S_{r_i}\}}\left(\alpha_{mr_i} I_{\{\tilde{D}_m > S_m\}}\right)\right) = \quad (9)$$

$$\alpha_{mr_i} P(D_{r_i} \leq S_{r_i}, \tilde{D}_m > S_m) = \alpha_{mr_i} P(D_{r_i} \leq S_{r_i}) - \alpha_{mr_i} P(\tilde{D}_{r_i} \leq S_{r_i}) P(\tilde{D}_m \leq S_m)$$

$$\frac{\partial^2 E(S_{r_i} - D_{r_i})^+}{\partial S_m^2} = \alpha_{mr_i}^2 f_{D_{r_i}|\tilde{D}_m > S_m}(S_{r_i}) P(\tilde{D}_m > S_m) - \alpha_{mr_i} f_{\tilde{D}_m}(S_m) P(\tilde{D}_{r_i} \leq S_{r_i})$$

$$> -\alpha_{mr_i} f_{\tilde{D}_m > S_m}(S_m) P(\tilde{D}_{r_i} \leq S_{r_i}) \quad (10)$$

$$\frac{\partial^2 \Pi_m(S_m, S_{r_1}, ..., S_{r_n})}{\partial S_m^2} < -(p_m - c + h_m) f_{D_m}(S_m) + \sum_{i=1}^{n} \alpha_{mr_i}(w_{r_i} - c) f_{\tilde{D}_m}(S_m) P(\tilde{D}_{r_i} \leq S_{r_i})$$

$$< -\left[(p_m - c + h_m) - \sum_{i=1}^{n} \alpha_{mr_i}(w_{r_i} - c)\right] f_{D_m}(S_m) \quad (11)$$

$$< 0$$

It is necessary to mention that the second inequality is obtained by $f_{\tilde{D}_m}(S_m) P(\tilde{D}_{r_i} \leq S_{r_i}) < f_{D_m}(S_m)$ and the third inequality is concluded from condition C1.

For investigating the submodularity of manufacturer profit function, it is sufficient to show that cross-partial derivatives are non-positive (Milgrom and Roberts, 1990). The cross partial derivative of $\Pi_m(S_m, S_{r_1}, ..., S_{r_n})$ is given as:

$$\frac{\partial^2 \Pi_m(S_m, S_{r_1}, ..., S_{r_n})}{\partial S_{r_i} \partial S_m} = -\sum_{\substack{j=1 \\ j \neq i}}^{n} \alpha_{mr_j} \alpha_{r_i r_j}(w_{r_j} - c) f_{D_{r_j}|(\tilde{D}_m > S_m, \tilde{D}_{r_i} > S_{r_i})}(S_{r_j}) P(\tilde{D}_m > S_m, \tilde{D}_{r_i} > S_{r_i})$$

$$- \alpha_{r_i m}(p_m - c + h_m) f_{D_m|\tilde{D}_{r_i} > S_{r_i}}(S_m) P(\tilde{D}_{r_i} > S_{r_i}) \quad (12)$$

$$- \alpha_{mr_i}(w_{r_i} - c) f_{D_{r_i}|\tilde{D}_m > S_m}(S_{r_i}) P(\tilde{D}_m > S_m)$$

which is obviously negative.

The profit function of retailer i coincides with the classical newsvendor problem with salvage value $(w_{r_i} - h_{r_i})$. The concavity of $\Pi_{r_i}(S_m, ..., S_{r_i}, ..., S_{r_n})$ with respect to its own decision variable $S_{r_i}$ is proved in Netessine and Rudi (2003). Similar to the previous section in order to prove the submodularity, cross-partial derivative of $\Pi_{r_i}(S_m,...,S_{r_i},...,S_{r_n})$ is obtained as follow:

$$\frac{\partial^2 \Pi_{r_i}(S_m,...,S_{r_i},...,S_{r_n})}{\partial S_j \partial S_{r_i}} = -\alpha_{jr_i}(p_{r_i} - w_{r_i} + h_{r_i})f_{D_{r_i}|\tilde{D}_j > S_j}(S_{r_i})P(\tilde{D}_j > S_j) < 0 \; (j \neq r_i) \quad (13)$$

which is clearly negative.

**Proof of Lemma 2.** Since the profit function of manufacturer and retailers are concave, there is a unique best response function for each channel which is obtained through first order conditions (Cachon and Netessine, 2004).

**Proof of preposition 1:** The sufficient conditions for existence of a pure Nash equilibrium are: (1) Strategy space for each player is nonempty, compact and convex subsets of a Euclidean space, (2) Payoff function for each player is continuous and (3) quasi-concave respect to her own decision variable (Nikaidô and Isoda, 1955). By choosing a large enough closed set $[0, \bar{S}_j]$ for each player, condition (1) is satisfied. In addition since the distribution function of demand is continuous, the continuity of players' payoff function on the strategy set is fulfilled. Also concavity of profit function of players is proved in lemma 1. Therefore there exists at least one pure Nash equilibrium.

In order to characterize the Nash equilibrium, it is sufficient to solve the system of best response functions. Since profit functions are concave, this system translates into the system of first order conditions.

**Proof of preposition 2:** If the strategy space of the game is convex and all equilibriums are interior, then $J(\frac{\partial \Pi_m}{\partial S_m}, \frac{\partial \Pi_{r_1}}{\partial S_{r_1}},..., \frac{\partial \Pi_{r_n}}{\partial S_{r_n}})$ is negative quasi-definite on the players' strategy set, equilibrium is unique (Friedman, 1990).

Jacobian $(\frac{\partial \Pi_m}{\partial S_m}, \frac{\partial \Pi_{r_1}}{\partial S_{r_1}},..., \frac{\partial \Pi_{r_n}}{\partial S_{r_n}})$ is given as:

$$H = \begin{bmatrix} \dfrac{\partial^2 \pi_m}{\partial S_m^2} & \dfrac{\partial^2 \pi_m}{\partial S_{r_1} \partial S_m} & \cdots & \dfrac{\partial^2 \pi_m}{\partial S_{r_n} \partial S_m} \\ \dfrac{\partial^2 \pi_{r_1}}{\partial S_m \partial S_{r_1}} & \dfrac{\partial^2 \pi_{r_1}}{\partial S_{r_1}^2} & \cdots & \dfrac{\partial^2 \pi_{r_1}}{\partial S_{r_n} \partial S_{r_1}} \\ \cdots & \cdots & \cdots & \cdots \\ \dfrac{\partial^2 \pi_{r_n}}{\partial S_m \partial S_{r_n}} & \dfrac{\partial^2 \pi_{r_n}}{\partial S_{r_1} \partial S_{r_n}} & \cdots & \dfrac{\partial^2 \pi_{r_n}}{\partial S_{r_n}^2} \end{bmatrix} \quad (14)$$

$H$ is negative quasi-definite if $H + H^T$ is negative definite (Friedman, 1990). Transpose of $H$ is as follow:

$$H^T = \begin{bmatrix} \dfrac{\partial^2 \pi_m}{\partial S_m^2} & \dfrac{\partial^2 \pi_{r_1}}{\partial S_m \partial S_{r_1}} & \cdots & \dfrac{\partial^2 \pi_{r_n}}{\partial S_m \partial S_{r_n}} \\ \dfrac{\partial^2 \pi_m}{\partial S_{r_1} \partial S_m} & \dfrac{\partial^2 \pi_{r_1}}{\partial S_{r_1}^2} & \cdots & \dfrac{\partial^2 \pi_{r_n}}{\partial S_{r_1} \partial S_{r_n}} \\ \cdots & \cdots & \cdots & \cdots \\ \dfrac{\partial^2 \pi_m}{\partial S_{r_n} \partial S_m} & \dfrac{\partial^2 \pi_{r_1}}{\partial S_{r_n} \partial S_{r_1}} & \cdots & \dfrac{\partial^2 \pi_{r_n}}{\partial S_{r_n}^2} \end{bmatrix} \quad (15)$$

and

$$H_T = H + H^T = \begin{bmatrix} \dfrac{2\partial^2 \pi_m}{\partial S_m^2} & \dfrac{\partial^2 \pi_{r_1}}{\partial S_m \partial S_{r_1}} + \dfrac{\partial^2 \pi_m}{\partial S_{r_1} \partial S_m} & \cdots & \dfrac{\partial^2 \pi_{r_n}}{\partial S_m \partial S_{r_n}} + \dfrac{\partial^2 \pi_m}{\partial S_{r_n} \partial S_m} \\ \dfrac{\partial^2 \pi_m}{\partial S_{r_1} \partial S_m} + \dfrac{\partial^2 \pi_{r_1}}{\partial S_m \partial S_{r_1}} & \dfrac{2\partial^2 \pi_{r_1}}{\partial S_{r_1}^2} & \cdots & \dfrac{\partial^2 \pi_{r_n}}{\partial S_{r_1} \partial S_{r_n}} + \dfrac{\partial^2 \pi_{r_1}}{\partial S_{r_n} \partial S_{r_1}} \\ \cdots & \cdots & \cdots & \cdots \\ \dfrac{\partial^2 \pi_m}{\partial S_{r_n} \partial S_m} + \dfrac{\partial^2 \pi_{r_n}}{\partial S_m \partial S_{r_n}} & \dfrac{\partial^2 \pi_{r_1}}{\partial S_{r_n} \partial S_{r_1}} + \dfrac{\partial^2 \pi_{r_n}}{\partial S_{r_1} \partial S_{r_n}} & \cdots & \dfrac{2\partial^2 \pi_{r_n}}{\partial S_{r_n}^2} \end{bmatrix} \quad (16)$$

According to Axes theorem since $H_T$ is symmetric, this matrix is negative definite if quadratic form of it on the players' strategy set is always negative (Poole, 2003). Quadratic form of $H_T$ is as follow:

$$Q(S) = S^T (H_T) S \quad (17)$$

$$Q(S_m, S_{r_1}, \ldots, S_{r_n}) = 2 * \sum_j \dfrac{\partial^2 \pi_j}{\partial S_j^2} S_j^2 + \sum_{\substack{k \\ k \neq j}} \sum_j \left( \dfrac{\partial^2 \pi_j}{\partial S_k \partial S_j} + \dfrac{\partial^2 \pi_k}{\partial S_j \partial S_k} \right) S_j S_k \; (\forall S_j, S_k \neq 0) \quad (18)$$

Consequently, according to concavity and submodularity features of payoff functions which is proved in lemma 1, quadratic form of $H_T$ on the interior point of strategy space is always negative. On the other hand, since $p_m - c > \sum_{i=1}^{n} \alpha_{mr_i}(w_{r_i} - c)$ at equilibrium point all of the channels stock the product and therefore equilibrium is interior point. Hence, it is concluded that under condition C2 equilibrium is unique.

**Appendix B: Stackelberg equilibrium**

**Proof of Lemma 3:** Since the best response function of retailer i is continuous, implicit function theorem is used to obtain the slope of it. By using this theorem, slope of best response function given as:

$$\frac{dS_{r_i}^{**}}{dS_m} = -\frac{\frac{\partial^2 \Pi_{r_i}}{\partial S_m \partial S_{r_i}}}{\frac{\partial^2 \Pi_{r_i}}{\partial S_{r_i}^2}} = -\frac{-\alpha_{mr_i}(p_{r_i} - w_{r_i} + h_{r_i}) f_{D_{r_i} | \tilde{D}_m > S_m}(S_{r_i}) p(\tilde{D}_m > S_m)}{-(p_{r_i} - w_{r_i} + h_{r_i}) f_{D_{r_i}}(S_{r_i})}$$

$$= -\frac{\alpha_{mr_i} f_{D_{r_i} | \tilde{D}_m > S_m}(S_{r_i}) p(\tilde{D}_m > S_m)}{f_{D_{r_i}}(S_{r_i})}$$

(19)

Therefore as shown in the expression, the slope of retailers' best response functions is always negative which implies that retailer i best response is monotonically decreasing in the manufacturer's strategy. Also as shown, the rate is less than substitution rate from the manufacturer to retailer i.

**Proof of Lemma 4:**

$$\frac{d}{dS_m}\left(\frac{\partial \Pi_m}{\partial S_{r_i}} \frac{dS_{r_i}^{**}}{dS_m}\right) = \frac{\partial^2 \Pi_m}{\partial S_{r_i}^2}\left(\frac{dS_{r_i}^{**}}{dS_m}\right)^2 + \frac{\partial \Pi_m}{\partial S_{r_i}} \frac{d^2 S_{r_i}^{**}}{dS_m^2} \tag{20}$$

By applying Base formula $\frac{dS_{r_i}^{**}}{dS_m}$ is given as:

$$\frac{dS_{r_i}^{**}}{dS_m} = -\frac{\alpha_{mr_i}\left(f_{D_{r_i}}(S_{r_i}) - f_{\tilde{D}_{r_i}}(S_{r_i}) P(\tilde{D}_m \leq S_m)\right)}{f_{D_{r_i}}(S_{r_i})}$$

$$= \left(\frac{\alpha_{mr_i} f_{\tilde{D}_{r_i}}(S_{r_i}) P(\tilde{D}_m \leq S_m)}{f_{D_{r_i}}(S_{r_i})} - \alpha_{mr_i}\right)$$

(21)

$$\frac{d^2 S_{r_i}^{**}}{dS_m^2} = -\frac{\alpha_{mr_i} f_{\tilde{D}_{r_i}}(S_{r_i}) f_{\tilde{D}_m}(S_m) f_{D_{r_i}}(S_{r_i}) - \alpha_{r_i m}^2 f'_{D_{r_i}|\tilde{D}_m > S_m}(S_{r_i}) P(\tilde{D}_m > S_m) f_{\tilde{D}_{r_i}}(S_{r_i}) P(\tilde{D}_m \leq S_m)}{f_{D_{r_i}}^2(S_{r_i})} \quad (22)$$

$$\frac{\partial^2 \Pi_m}{\partial S_{r_i}^2}\left(\frac{dS_{r_i}^{**}}{dS_m}\right)^2 + \frac{\partial \Pi_m}{\partial S_{r_i}} \frac{d^2 S_{r_i}^{**}}{dS_m^2} = [-(p_m - c + h_m)\left(\alpha_{r_i m}^2 f_{D_{r_i}|\tilde{D}_{r_i} > S_{r_i}}(S_m) P(\tilde{D}_{r_i} > S_{r_i}) - \alpha_{r_i m} f_{\tilde{D}_{r_i}}(S_{r_i}) P(\tilde{D}_m \leq S_m)\right) -$$
$$\sum_{\substack{l=1 \\ l \neq i}}^{l=n}(w_{r_i} - c)\left(\alpha_{r_i r_i}^2 f_{D_{r_i}|\tilde{D}_{r_i} > S_{r_i}}(S_{r_i}) P(\tilde{D}_{r_i} > S_{r_i}) - \alpha_{r_i r_i} f_{\tilde{D}_{r_i}}(S_{r_i}) P(\tilde{D}_{r_i} \leq S_{r_i})\right) -$$
$$\left((w_{r_i} - c) f_{D_{r_i}}(S_{r_i})\right)] * \left(\frac{\alpha_{mr_i} f_{D_{r_i}|\tilde{D}_m > S_m}(S_{r_i}) P(\tilde{D}_m > S_m)}{f_{D_{r_i}}(S_{r_i})}\right)^2 +$$
$$[-\alpha_{r_i m}(p_m - c + h_m) P(D_m \leq S_m | \tilde{D}_{r_i} > S_{r_i}) P(\tilde{D}_{r_i} > S_{r_i}) -$$
$$\sum_{\substack{l=1 \\ l \neq i}}^{l=n} \alpha_{r_i r_i}(w_{r_i} - c) P(D_{r_i} \leq S_{r_i} | \tilde{D}_{r_i} > S_{r_i}) P(\tilde{D}_{r_i} > S_{r_i}) +$$
$$\left((w_{r_i} - c) P(D_{r_i} > S_{r_i})\right)] * [\left(\frac{\alpha_{mr_i} f_{\tilde{D}_{r_i}}(S_{r_i}) f_{\tilde{D}_m}(S_m) f_{D_{r_i}}(S_{r_i})}{f_{D_{r_i}}^2(S_{r_i})}\right) -$$
$$\left(\frac{\alpha_{mr_i}^2 f'_{D_{r_i}|\tilde{D}_m > S_m}(S_{r_i}) f_{\tilde{D}_{r_i}}(S_{r_i}) P(\tilde{D}_m \leq S_m) P(\tilde{D}_m > S_m)}{f_{D_{r_i}}^2(S_{r_i})}\right)] \quad (23)$$

$$\frac{\partial^2 \Pi_m}{\partial S_{r_i}^2}\left(\frac{dS_{r_i}^{**}}{dS_m}\right)^2 + \frac{\partial \Pi_m}{\partial S_{r_i}} \frac{d^2 S_{r_i}^{**}}{dS_m^2} < [\left(\alpha_{r_i m} \alpha_{mr_i}^2 (p_m - c + h_m) f_{\tilde{D}_{r_i}}(S_{r_i}) f_{D_{r_i}|\tilde{D}_m > S_m}^2(S_{r_i}) P(\tilde{D}_m \leq S_m) P^2(\tilde{D}_m > S_m)\right) +$$
$$\left(\sum_{\substack{l=1 \\ l \neq i}}^{l=n} \alpha_{r_i r_i} \alpha_{mr_i}^2 (w_{r_i} - c) f_{\tilde{D}_{r_i}}(S_{r_i}) f_{D_{r_i}|\tilde{D}_m > S_m}^2(S_{r_i}) P^2(\tilde{D}_m > S_m) P(\tilde{D}_{r_i} \leq S_{r_i})\right) -$$
$$\left(\alpha_{mr_i}^2 (w_{r_i} - c) f_{D_{r_i}}(S_{r_i}) f_{D_{r_i}|\tilde{D}_m > S_m}^2(S_{r_i}) P^2(\tilde{D}_m > S_m)\right)] / [f_{D_{r_i}}^2(S_{r_i})] +$$
$$[-\alpha_{r_i m}(p_m - c + h_m) P(D_m \leq S_m | \tilde{D}_{r_i} > S_{r_i}) P(\tilde{D}_{r_i} > S_{r_i}) -$$
$$\sum_{\substack{l=1 \\ l \neq i}}^{l=n} \alpha_{r_i r_i}(w_{r_i} - c) P(D_{r_i} \leq S_{r_i} | \tilde{D}_{r_i} > S_{r_i}) P(\tilde{D}_{r_i} > S_{r_i}) +$$
$$\left((w_{r_i} - c) P(D_{r_i} > S_{r_i})\right)] * [\left(\frac{\alpha_{mr_i} f_{\tilde{D}_{r_i}}(S_{r_i}) f_{\tilde{D}_m}(S_m) f_{D_{r_i}}(S_{r_i})}{f_{D_{r_i}}^2(S_{r_i})}\right) -$$
$$\left(\frac{\alpha_{mr_i}^2 f'_{D_{r_i}|\tilde{D}_m > S_m}(S_{r_i}) f_{\tilde{D}_{r_i}}(S_{r_i}) P(\tilde{D}_m \leq S_m) P(\tilde{D}_m > S_m)}{f_{D_{r_i}}^2(S_{r_i})}\right)] \quad (24)$$

By considering $A_i$ and $B_i$ as bellow:

$$A_i = \alpha_{mr_i}^2 f_{D_{r_i}|\tilde{D}_m > S_m}^2(S_{r_i}) P^2(\tilde{D}_m > S_m) f_{D_{r_i}}(S_{r_i}) \quad (25)$$

$$B_i = \alpha_{mr_i}^2 f'_{D_{r_i}|\tilde{D}_m > S_m}(S_{r_i}) f_{\tilde{D}_{r_i}}(S_{r_i}) P(\tilde{D}_m \leq S_m) P(\tilde{D}_m > S_m) P(D_{r_i} > S_{r_i}) -$$
$$\alpha_{mr_i} f_{\tilde{D}_{r_i}}(S_{r_i}) f_{\tilde{D}_m}(S_m) f_{D_{r_i}}(S_{r_i}) P(D_{r_i} > S_{r_i}) \quad (26)$$

it is concluded that:

$$\frac{\partial^2 \Pi_m}{\partial S_{r_i}^2}\left(\frac{dS_{r_i}^{**}}{dS_m}\right)^2 + \frac{\partial \Pi_m}{\partial S_{r_i}}\frac{d^2 S_{r_i}^{**}}{dS_m^2} < \frac{\alpha_{r_i m}(p_m - c + h_m)A_i + \sum_{\substack{l=1\\l\neq i}}^{n}\alpha_{r_i r_l}(w_{r_l} - c)A_i - (w_{r_i} - c)A_i}{f_{D_{r_i}}^2(S_{r_i})} +$$

$$\frac{\alpha_{r_i m}(p_m - c + h_m)P(D_m \leq S_m | \tilde{D}_{r_i} > S_{r_i})B_i}{f_{D_{r_i}}^2(S_{r_i})} + \qquad (27)$$

$$\frac{\sum_{\substack{l=1\\l\neq i}}^{n}\alpha_{r_i r_l}(w_{r_l} - c)P(D_{r_l} \leq S_{r_l} | \tilde{D}_{r_i} > S_{r_i})B_i - (w_{r_i} - c)B_i}{f_{D_{r_i}}^2(S_{r_i})}$$

$$< \frac{-(A_i + B_i)}{f_{D_{r_i}}^2(S_{r_i})}\left[(w_{r_i} - c) - \alpha_{r_i m}(p_m - c + h_m) - \sum_{\substack{l=1\\l\neq i}}^{n}\alpha_{r_i r_l}(w_{r_l} - c)\right]$$

This inequality is obtained through following expression:

$$f_{\tilde{D}_j}(S_j)P(\tilde{D}_k \leq S_k) < f_{D_j}(S_j) \quad (j,k = M, R_1, \ldots, R_n, j \neq k) \qquad (28)$$

$$P(D_{r_i} > S_{r_i}) \geq P(\tilde{D}_{r_i} > S_{r_i}) \qquad (29)$$

Therefore under condition C3, $\frac{d}{dS_m}\left(\frac{\partial \Pi_m}{\partial S_{r_i}}\frac{dS_{r_i}^{**}}{dS_m}\right) \leq 0$ if:

$$A_i + B_i = \alpha_{mr_i}^2 f_{D_{r_i}|\tilde{D}_m > S_m}^2(S_{r_i})P^2(\tilde{D}_m > S_m)f_{D_{r_i}}(S_{r_i})$$

$$\alpha_{mr_i}^2 f'_{D_{r_i}|\tilde{D}_m > S_m}(S_{r_i})f_{\tilde{D}_{r_i}}(S_{r_i})P(\tilde{D}_m \leq S_m)P(\tilde{D}_m > S_m)P(D_{r_i} > S_{r_i}) - \qquad (30)$$

$$\alpha_{mr_i} f_{\tilde{D}_{r_i}}(S_{r_i})f_{\tilde{D}_m}(S_m)f_{D_{r_i}}(S_{r_i})P(D_{r_i} > S_{r_i}) \geq 0$$

**Proof of preposition 3:** In order to prove uniqueness of Stackelberg equilibrium, it is sufficient to show that $\Pi_m\left(S_m, S_{r_1}^{**}, \ldots, S_{r_n}^{**}\right)$ with respect to its own decision variable $S_m$ is quasi concave (Cachon and Netessine, 2004).

$$\frac{d^2\Pi_m}{dS_m^2} = \frac{d}{dS_m}\left(\frac{\partial \Pi_m}{\partial S_m} + \frac{\partial \Pi_m}{\partial S_{r_1}}\frac{dS_{r_1}^{**}}{dS_m} + \ldots + \frac{\partial \Pi_m}{\partial S_{r_n}}\frac{dS_{r_n}^{**}}{dS_m}\right)$$

$$= \frac{\partial^2 \Pi_m}{\partial S_m^2} + \sum_{i=1}^{i=n}\frac{\partial^2 \Pi_m}{\partial S_{r_i}^2}\left(\frac{dS_{r_i}^{**}}{dS_m}\right)^2 + \frac{\partial \Pi_m}{\partial S_{r_i}}\frac{d^2 S_{r_i}^{**}}{dS_m^2} \qquad (31)$$

According to Lemma 1 fulfillment of condition C1 is sufficient to have $\frac{\partial^2 \Pi_m}{\partial S_m^2} < 0$. Also if condition C3 and sufficient condition of Lemma 4 is satisfied, $\frac{\partial^2 \Pi_m}{\partial S_{r_i}^2}\left(\frac{dS_{r_i}^{**}}{dS_m}\right)^2 + \frac{\partial \Pi_m}{\partial S_{r_i}}\frac{d^2 S_{r_i}^{**}}{dS_m^2} \leq 0$ and therefore $\frac{d^2 \Pi_m}{dS_m^2} < 0$.

**Proof of preposition 4:**

$$\frac{d\Pi_m(S_m, S_{r_1}^{**}, ..., S_{r_n}^{**})}{dS_m} = \frac{\partial \Pi_m}{\partial S_m} + \frac{\partial \Pi_m}{\partial S_{r_1}}\frac{dS_{r_1}^{**}}{dS_m} + ... + \frac{\partial \Pi_m}{\partial S_{r_n}}\frac{dS_{r_n}^{**}}{dS_m} = 0 \quad (32)$$

$$\frac{\partial \Pi_m}{\partial S_m} = (p_m - c) - (p_m - c + h_m)P(D_m \leq S_m) - \sum_{i=1}^{i=n} \alpha_{mr_i}(w_{r_i} - c)P(D_{r_i} \leq S_{r_i}, \tilde{D}_m > S_m) \quad (33)$$

$$\frac{\partial \Pi_m}{\partial S_{r_i}} = -\alpha_{r_i m}(p_m - c + h_m)P(D_m \leq S_m, \tilde{D}_{r_i} > S_{r_i}) + (w_{r_i} - c)P(D_{r_i} > S_{r_i}) - \sum_{\substack{l=1 \\ l \neq i}}^{l=n} \alpha_{r_i r_l}(w_{r_l} - c)P(D_{r_l} \leq S_{r_l}, \tilde{D}_{r_i} > S_{r_i}) \quad (34)$$

If $(S_m^*, S_{r_1}^*, ..., S_{r_n}^*)$ is the Nash equilibrium of simultaneous game, at this point $\frac{\partial \Pi_m}{\partial S_m} = 0$. On the other hand respect to the assumption $w_{r_i} = c$ $(i = 1,...,n)$, $\frac{\partial \Pi_m}{\partial S_{r_i}} < 0$ and hence since $\frac{dS_{r_i}}{dS_m} < 0$, it is concluded that $\frac{d\Pi_m}{dS_m} > 0$. Therefore it is concluded that the manufacturer carries more inventory than a simultaneous move game.